\input{aipcheck.tex}

  \def\selectedoptions{final}
\documentclass[
   \selectedoptions
  ]
  {aipproc}

\def\selectedlayoutstyle{8x11single} \layoutstyle\selectedlayoutstyle

\SetInternalRegister\hbadness{8000} % pseudo latin isn't breaking very well :-)

\newcommand\doingARLO[2][]{%
  \ifx\mmref\undefined #1\else #2\fi
}

\begin{document}

\title
      {Homologous Core Collapse in a Massive Star and Self-Similar Evolution of Rebound Shocks }


\author{Yi Cao}{
  address={Department of Physics and Tsinghua Center for Astrophysics,
  Tsinghua University, Beijing, 100084, China},
  email={y-cao04@mails.thu.edu.cn},
  thanks={This work was commissioned by the AIP}
}
\iftrue
\author{Yu-Qing Lou}{
  address={Department of Physics and Tsinghua Center for Astrophysics,
  Tsinghua University, Beijing, 100084, China},
  email={louyq@mail.tsinghua.edu.cn},
}

\fi
% August     12,      Tuesday       Kiel morning, noon
% August     15,      Friday        Kiel noon, afternoon, evening
%                                   Institut linux machine
% August     16,      Saturday      Kiel swimming in the Baltic sea
%                                   afternoon, evening
% August     17,      Sunday        Kiel morning, noon, afternoon
%                                   evening, night
% August     19,      Tuesday       Kiel morning, noon, afternoon
%                                   evening
% August     25,      Monday        Kiel morning, noon, afternoon
%
% \copyrightholder{Acoustical Scociety of America}
% \copyrightyear  {2001}

\begin{abstract}
During the gravitational core collapse of a massive progenitor
star which may give rise to at least a class of gamma-ray bursts
(GRBs) associated with supernovae, a stellar core rapidly passes
through a short yet important phase of neutronization, producing a
huge amount of energetic neutrinos and photons which contribute to
the total pressure within the progenitor core. The collection of
neutrinos, photons and gas materials together may be approximated
as a fluid with a polytropic index $\gamma=4/3$ under the action
of self-gravity. With a substantial generalization and using
analytical and numerical methods (Lou \& Cao 2008), we recently
constructed and examined various self-similar solutions to
describe collapses, rebound shocks and flows systematically in a
$\gamma=4/3$ polytropic gas mixture with spherical symmetry, and
compare our results with those of Goldreich \& Weber (1980). It is
also possible to construct central void solutions without or with
shocks. Various features and characteristics of this nonlinear
relativistically hot gas dynamics, including asymptotic and exact
solutions, are presented. This more general polytropic model
analysis provides the dynamic basis of understanding the evolution
of rebound shocks in supernovae (SNs) and the results may be also
utilized to benchmark hydrodynamic simulations.
\end{abstract}

\date{\today}
\maketitle

\section{Introduction}

\ifthenelse{\equal\selectedlayoutstyle{6x9}}{\par\bfseries
  Note: The entire paper will be reduced 15\% in the printing
  process. Please make sure all figures as well as the text within the
  figures are large enough in the manuscript to be readable in the
  finished book.\par\bfseries
  Note: The entire paper will be reduced 15\% in the printing
  process. Please make sure all figures as well as the text within the
  figures are large enough in the manuscript to be readable in the
  finished book.\par\bfseries
  Note: The entire paper will be reduced 15\% in the printing
  process. Please make sure all figures as well as the text within the
  figures are large enough in the manuscript to be readable in the
  finished book.\normalfont}{}

Long gamma-ray bursts (GRBs) are generally thought to originate
from demises and explosions of massive stars. They release the
gravitational energy by drastic core collapses and shock
outbursts, producing characteristic radiations in different energy
bands, including gamma rays, X-rays and energetic neutrinos. While
the key mechanism of GRBs still remains mysterious, various
observations offer valuable clues to these extremely explosive
``cosmic fireworks''. Especially since the discovery of the
precise match in positions between GRB980425 and SN1998bw, several
GRBs are clearly associated with at least one class of supernovae
(Type Ibc) which are related to deaths of old massive stars. Three
other such associations GRB030329 and SN2003dh, GRB031203 and
SN2003lw, GRB060218 and SN2006aj together with a recent X-Ray
Outburst (XRO)-SN connection between XRO 080109 and SN 2008D
(e.g., Woosley \& Bloom 2006; Soderberg 2008) were reported within
the last decade.
%} {\bf Other three association examples and references as well
%as SN2008D.} In addition, several similar events including an
%association between X-Ray Flare (XRF) and SN were
%observed. {\bf References} As GRBs, espectially
For long GRBs,
%and SNe, seems closely associated in several
%cases {\it (Woosley \& Bloom 2006)},
dynamics of stellar core collapses and rebound shock evolution in
the context of SN explosions are important to both theories and
observations of GRBs.

With certain assumptions about the state of collapsing stellar
cores, hydrodynamics and magnetohydrodynamics (MHD) may be adopted
to model these stellar collapses with or without shocks (e.g., Lou
\& Wang 2006, 2007). Based on gas properties, a self-similar
dynamics may emerge in these astrophysical processes. First, to
simplify the problem, a quasi-spherical symmetry is presumed and
only self-gravity and thermal pressure are taken into account in
this semi-analytic model of a relativistically hot polytropic gas.
%A key to this model is the assumption to the state of the star.
A simple yet useful assumption is the conventional polytropic
state function, i.e., $P=\kappa\rho^\gamma$ where $P$ and $\rho$
are pressure and density, $\gamma$ is the polytropic index and
$\kappa$ is a global constant. Based on such an assumption,
valuable information about the stellar structure, solar physics,
star formation and SN explosions may be obtained (e.g.,
Chandrasekhar 1939; Shu 1977; Lou \& Shen 2004; Suto \& Silk 1988;
Lou \& Wang 2006, 2007). A simple argument on total energy of a
static polytropic sphere demonstrates that $\gamma>4/3$ cases are
stable while $\gamma<4/3$ ones are unstable (Chandrasekhar 1939).
Thus most researchers focus on
%Therefore, researchers pay most attentions to
$1\le\gamma<4/3$ cases.

We emphasize that the critical case of $\gamma=4/3$ is important
in astrophysical contexts. Statistical physics illustrates that in
a temperature much lower than the Fermi energy, the state of
particles whose rest mass is negligible is a $\gamma=4/3$
polytrope. Meanwhile, numerical simulations show that $\gamma=4/3$
polytrope is a sensible approximation for the central state of a
compact object. In addition, the pressures of neutrinos
%which had been collected from SN1987,
and of photons, both crucial in collapsing stellar cores, may also be
approximated by a $\gamma=4/3$ polytropic state.
%Nevertheless, in the framework of conventional polytropic assumption,
%few researches concerned this special case systematically. After we
%deal with this case systematically in a relatively general situation,
It turns out that the conventional polytropic assumption limits
$\gamma=4/3$ situation severely in a self-similar model. In a
classic paper on $\gamma=4/3$, Goldreich \& Weber (1980; GW
hereafter) considered homologous collapse and concluded that when
the pressure decreases by a fraction of no more than $2.9\%$, a
homologous core collapse would occur in the stellar interior,
which is considerably lower than a value of $26\%$ from numerical
simulations of Bethe et al. (1979). GW tried to reduce this
difference by introducing an inner core in a progenitor. Yahil
(1983) performed a polytropic analysis and treated GW results as a
limit of $\gamma\rightarrow (4/3)^-$.

In contrast to the conventional gas of a constant $\kappa$, a more
general polytropic gas still obeys
\begin{eqnarray}
\frac{D}{Dt}\log\left(\frac{P}{\rho^\gamma}\right)=\left(\frac{\partial}{\partial
t}+\vec{u}\cdot\nabla\right)\log\left(\frac{P}{\rho^\gamma}\right)=0\
\label{eqst}
\end{eqnarray}
for the specific entropy conservation along streamlines with a
variable $P/\rho^\gamma$
%is also utilized in previous researches (for example,
(e.g., Cheng 1977, 1978).
%Physically, for a polytropic gas, $\log(P/\rho^\gamma)$
%represents local specific entropy and the operator $D/Dt$
%is to derivative along the streamline. Hence, the state
%assumption means conservation of local specific entropy.
Clearly, the conventional polytrope is only a special case.
Nonetheless, mathematical derivations become more involved if the
more general polytrope is adopted in the model analysis.
%Fortunately Fatuzzo et al. (2004) provided a new type of
%self-similar transformation to solve this mathematical trouble.
%The focus in their paper is the nonzero velocity at infinity.

Recently, Lou \& Cao (2008) used the self-similar transformation
(Fatuzzo et al. 2004)
%'s transformation to particularly concern
for a more general polytrope of $\gamma=4/3$ (see eqn
(\ref{eqst})). Based on different values of a scaling index $a$,
the problem is divided into three classes.
%to discuss separately. A systematic analysis is performed and
%results are presented below in order.
Some essential procedures, analyses, and results are summarized in
the following.

\section{Model formulation and main results}

In spherical polar coordinates ($r$, $\theta$, $\phi$) with
spherical symmetry, physical variables are functions of radius $r$
and time $t$. Conservations of mass and momentum together with the
local specific entropy conservation along streamlines govern the
dynamic evolution of a relativistically hot gas of $\gamma=4/3$.
%which give a set of partial differential equations (PDEs).
Without shocks, the problem is invariant under the time reversal
transformation. Thus, an expansion solution can also describe a
collapse solution by such a transformation. In order to solve
hydrodynamic partial differential equations (PDEs), a self-similar
transformation is introduced to reduce PDEs to ordinary
differential equations (ODEs), namely
\begin{eqnarray}
x=At^ar\ ,\qquad\rho=\frac{\alpha(x)}{4\pi Gt^2}\ ,\qquad
M=\frac{m(x)}{A^3Gt^{3a+2}}\ ,\qquad u=\frac{v(x)}{At^{a+1}}\
,\qquad P=\frac{p(x)}{4\pi GA^2 t^{2(a+2)}}\ ,
\end{eqnarray}
where $\rho$, $M$, $u$ and $P$ are respectively density, enclosed
mass, radial velocity and pressure, $a$ is a scaling index in such
a transformation, $G$ is the gravitational constant and $A$ is a
dimensional parameter to make the independent self-similar
variable $x$ dimensionless. Then the governing hydrodynamic PDEs
are reduced to nonlinear ODEs
\begin{eqnarray}
(ax+v)x^2\alpha=(3a+2)m\ ,\label{eq1}\\
%\frac{dm}{dx}=x^2\alpha\ ,\label{eq2}\\
(ax+v)\frac{dv}{dx}+\frac{1}{\alpha}\frac{dp}{dx}
=-\frac{m}{x^2}+(a+1)v\ ,\label{eq3}\\
(ax+v)\frac{d}{dx}\log\bigg(\frac{p}{\alpha^\gamma}\bigg)
=2(2+a-\gamma)\ ,\label{eq4}\\
(ax+v)\frac{1}{\alpha}\frac{d\alpha}{dx}+\frac{dv}{dx}
=2\left(1-\frac{v}{x}\right)\ .\label{eq5}
\end{eqnarray}
For considerations of dynamic processes in the core of a massive
star and especially rebound shocks of SNe associated with GRBs, we
set $\gamma=4/3$ to approximate the state of gas materials in the
compact central core. Generally speaking, we require $a$ being
less than zero. Importantly, equation (\ref{eq1}) implies a
division of all cases into three classes, depending on the value
of $a$ being greater than, equal to or less than $-2/3$. In the
following analysis, we deal with these classes separately and
obtain respective results. Comparing to GW, our analysis to the
$a=-2/3$ class represents a substantial generalization.

%\section{analysis and main results}
%In this section, three separate classes will be discussed
%individually.
\paragraph{The case of $a=-2/3$}
For $a=-2/3$, the reduced specific entropy conservation
(\ref{eq4}) becomes automatically satisfied. Meanwhile from
equations (\ref{eq1})$-$(\ref{eq5}), one can readily derive the
following two relations
\begin{eqnarray}
%v=-\frac{2}{3}x
v=\frac{2}{3}x\ ,\qquad\qquad\qquad
\frac{1}{x^2}\frac{d}{dx}\left(\frac{x^2}{\alpha}
\frac{dp}{dx}\right)=-\alpha+\frac{2}{3}\ .\label{eq6}
\end{eqnarray}
It is taken that the pressure $P$ is proportional to $\rho^{4/3}$,
and we need to know the proportional coefficient as a function of
$(r,\ t)$ to complete the problem. Physically,
$\log(P/\rho^{4/3})$ is proportional to specific entropy $s(r,\
t)$ in a polytropic gas. Once this distribution $s(r,\ t)$ is
known, the gas dynamics is determined. A more general case in
self-similar transformation is to allow $p=g(x)\alpha^{4/3}$ where
$g(x)$ is a function of $x$ to be specified for a self-similar
flow. For a given $g(x)$, equation (\ref{eq6}) can be solved with
proper initial or boundary conditions. In general, velocity cannot
diverge at large radii. Therefore a boundary condition to this
problem is that an outer boundary of zero density should exist.
% so that the whole system is cut off there.
%{\bf Stop checking here}

The first cut is to adopt a constant specific entropy $s(r,\ t)$
everywhere, i.e., $g(x)=1$, as GW did. We confirmed the main
results of GW. An essential result is that the central density has
a critical minimum below which the solution of equation
(\ref{eq6}) has no vanishing $\alpha$ at a finite radius.
%, which violets the outer boundary condition.
This critical value of $\alpha$
%{\bf You mean $\alpha$?}
is $\sim 101.88$, consistent with GW work.
%When applying this to astrophysics,
In the context of a stellar collapse, GW result is somewhat
limited in certain aspects.
%meets a severe limitation.
Take an old star in a relatively stationary state as described by
the Lane-Emden equation (e.g., Chandrasekhar 1939). As the central
nuclear burning becomes insufficient,
%radiation and neutrino pressures in the center against
%the gravitational force becomes weaker and weaker.
the whole system rapidly evolves into a homologous collapsing or
pre-collapsing phase. GW estimated the largest pressure reduction
fraction (denoted by $r_c$ here) for a homologous collapse (prior
to a rebound)
%(which we denote by $r_c$)
and obtained a value of $\sim 2.9\%$ under their assumption for a
polytropic sphere of $\gamma=4/3$. In contrast, numerical
simulations of Bethe et al. (1979) show that this value $r_c$ in
supernovae could be substantially larger, about $26\%$. To
reconcile, GW proposed that there is an inner and smaller core
which obeys their results as the pre-collapsing phase.
%However, no observation evidence bolsters such an explanation
%and the assumption of a constant specific entropy distribution
%is too idealized so it may deviate a lot from real situations.

To generalize GW analysis for a homologous collapse, we allow a
fairly arbitrary form of $g(x)$ in our model and thus accommodate
a broad class of solutions for the density profile. Such a
variable $s(r,\ t)$
%non-constant distribution of specific entropy
may lead to a consistent result with numerical simulations (e.g.,
Bethe et al. 1979).
%Although specific entropy is not an observable
%quantity so that the actual entropy distribution remains unknown,
%some kind of forms are manually given to $g(x)$ as a test.
In our numerical exploration, we choose $g(x)=1/(1+\varepsilon x)$
with $\varepsilon$ being a coefficient to measure the range of
$g(x)$ variation. According to our experiments, when
$\varepsilon=0.1$, the pressure reduction $r_c$ can reach $\sim
10\%$ and when $\varepsilon=0.3$, $r_c$ is $\sim 26\%$ as given by
Bethe et al. (1979).
%Though it is entirely possible that this consistency is occasional, t
Given idealizations, these explorations do reveal a simple fact
that a variable distribution of specific entropy $s(r,\ t)$ can be
extremely important and we could model this effect in a
self-similar manner.
%may perform an important role, which can
%be included in our self-similar model.

\paragraph{The case of $a<-2/3$}
In these cases, equations (\ref{eq1}) and (\ref{eq4}) lead to a
new form of state function below
%{\bf Please pay attention to notational consistency!}
\begin{eqnarray}
p=C_0m^q\alpha^\gamma\ ,\qquad\qquad\qquad
q\equiv\frac{2(2+a-\gamma)}{(3a+2)}\ ,
\end{eqnarray}
where $C_0$ is a constant.
%This state function has an obvious but profound physical meaning.
Most importantly, the variable specific entropy
$\log(P/\rho^\gamma)$ is connected with the enclosed mass $m(x)$
in a power-law form.
%according to such a state function.
The conservation of mass guarantees the conservation of specific
entropy. For $q=0$ and thus $a=\gamma-2$, we have a conventional
polytrope; if we further require $\gamma=1$ and thus $a=-1$, we
then have an isothermal gas (Shu 1977; Lou \& Shen 2004). For
$\gamma=4/3$ under consideration, $q=2/3$, i.e.,
%$p=C_0m^{2/3}\rho^{4/3}$, which is a completely different family
$p=C_0m^{2/3}\alpha^{4/3}$, which is a completely different family
of problem in contrast to $P=\kappa\rho^{4/3}$. Note that for
$\gamma\neq 4/3$ cases, the constant $C_0$ can be absorbed by
readjusting the $A$ parameter in the transformation but for
$\gamma=4/3$, constant $C_0$ is independent and characterizes the
local sound speed.

It is easy to derive two coupled nonlinear ODEs for $v(x)$ and
$\alpha(x)$ from equations (\ref{eq1})$-$(\ref{eq5}).
%\begin{eqnarray}
%(ax+v)\frac{dv}{dx}+\frac{4C_0}{3}x^{4/3}
%\left(\frac{ax+v}{3a+2}\right)^{2/3} \frac{d\alpha}{dx}=
%-\frac{ax+v}{3a+2}\alpha+(a+1)v-\frac{2C_0}{3}
%\left(\frac{ax+v}{3a+2}\right)^{-1/3}x^{4/3}\alpha\  ,\label{eq11}\\
%\frac{dv}{dx}+\frac{(ax+v)}{\alpha}\frac{d\alpha}{dx}
%=2\left(1-\frac{v}{x}\right)\ \qquad.\label{eq12}
%\end{eqnarray}
Through analysis, we obtain all kinds of relevant solutions both
asymptotically and numerically, with two globally analytical
solutions. Numerical simulations are needed to decide the paths of
various self-similar evolution.

In our analysis, we compare with known results, finding more
general counterparts of known situations, such as the central
free-fall solution (Larson 1969; Penston 1969), the expansion wave
collapse solution (EWCS; Shu 1977), envelope expansion with core
collapse solutions (EECC; Lou \& Shen 2004) and quasi-static
solutions (Lou \& Wang 2006, 2007). Furthermore, two new types of
asymptotic solutions are also found. Corresponding physical
meanings are discussed for possible processes in astrophysics. To
enlarge the range of astrophysical applications, shock waves are
also included. For shock conditions, mass, momentum and energy
should be all conserved across a shock front. Numerical solutions
connecting different types of asymptotic solutions are constructed
in our model, with relevant physical scenarios outlined (see Lou
\& Cao 2008 for details).
%For detailed analysis and calculations,
%please refer to Lou \& Cao (2008).

Most interestingly, we can construct central void solutions
without or with shocks in the framework of hydrodynamics involving
thermal pressure and self-gravity. In a background of Einstein de
Sitter expanding universe, Fillmore \& Goldreich (1984a, b)
considered a collection of collisionless particles with
self-gravity. In the presence of certain kinds of perturbations,
voids can be generated surrounding the center of such a flow of
collisionless particles. Similarly in our model, although without
an expanding background, thermal pressure also has a tendency of
expansion and thus provides a driving mechanism to produce
expanding voids under favorable conditions.
%In practice, in
%equation (\ref{eq1}), the reduced enclosed mass $m$ vanishes as long
%as $v=-ax$. A solution starts from a point in the line $v=-ax$
%instead of the origin, and goes outwards, either smoothly or with
%weak discontinuity or shock waves. Then a void solution is constructed.
Observationally, voids are ubiquitous in various astrophysical
contexts, such as supernova remnants, molecular clouds, hot
bubbles, superbubbles, planetary nebulae and galaxy clusters (see
Hu \& Lou 2008 and Lou \& Hu 2008 for more details).

\paragraph{The case of $a>-2/3$}
Since the enclosed mass cannot be negative, $a>-2/3$ requires
$v>-ax$ in equation (\ref{eq1}). However, as we cannot obtain any
solution with an outer boundary of finite radius, velocity should
converge as $x$ goes to infinity. However, these two requirements
are incompatible with each other. The requirement of $v>-ax$ will
certainly lead to a divergent velocity at large $x$. These cases
are thus excluded in our model consideration.

\section{Summary and conclusions}

%To sum up, we present some of main contexts
%of Lou \& Cao (2008) in this paper.
Both analytic and numerical self-similar solutions have been
explored for a more general polytropic gas with $\gamma=4/3$ (Lou
\& Cao 2008; Wang \& Lou 2008). We have directly extended the
classical analysis of GW for a conventional polytropic gas of
spherical symmetry with $\kappa=1$, giving a plausible resolution
to the discrepancy between theoretical calculations (GW) and
numerical simulations (Bethe et al. 1979). We then study
distributions of specific entropy in a certain system or infer
these from other available information on the basis of our
theoretical knowledge. Once such an entropy distribution is known,
the self-similar dynamical evolution of a homologous core collapse
can be calculated readily from our model. Another family of
solutions are also discussed with a new state function of
$P\propto M^{2/3}\rho^{4/3}$. In addition to counterparts of
various previously know types of analytic and asymptotic solutions
at $\gamma\neq 4/3$, we find some new ones and discuss their
physical implications. In Lou \& Cao (2008), we attempt to analyze
the self-similar dynamics of a $\gamma=4/3$ polytropic gas
systematically and hope that it would provide valuable and useful
clues to the study of compact objects and high-energy explosive
phenomena, such as shock breakouts of SN explosions and thus GRBs.

There are several physical processes, such as nuclear reactions,
radiation pressure, neutrino transportation, general relativity,
rotation and magnetic field effects, that are not included in our
preliminary model analysis.
%which may actually be vital for some
%observational results, such as relativistic jets.
However, given approximations and idealizations of our model, we
hope that the results of this theoretical hydrodynamic analysis
catch certain essential features of large-scale shocks in a
relativistically hot gas.

\begin{theacknowledgments}
This research has been supported in part by the National Natural
Science Foundation of China (NSFC) grants 10373009 and 10533020 at
Tsinghua University, by the SRFDP 20050003088, the Yangtze
Endowment and the National Undergraduate Innovation Training
Project from the Ministry of Education at Tsinghua University, and
by Tsinghua Center for Astrophysics (THCA).
%, and by the ASCI Center for Astrophysical Thermonuclear
%Flashes at the University of Chicago.
\end{theacknowledgments}

% choose bibtex style depending on layout style and options used in
% sample:

\doingARLO[\bibliographystyle{aipproc}]
          {\ifthenelse{\equal{\AIPcitestyleselect}{num}}
             {\bibliographystyle{arlonum}}
             {\bibliographystyle{arlobib}}
          }
%\bibliography{CaoLou}

\end{document}